\documentclass[a4paper,12pt]{article}
\usepackage{CJK}
\usepackage{color}
\usepackage{graphicx}
\usepackage{epstopdf}
\usepackage{amssymb}
\usepackage{amsfonts}
\usepackage{amsmath}
\usepackage{mathrsfs}
\usepackage{ulem}
\usepackage[numbers,sort&compress]{natbib}
\usepackage{epsf}
\usepackage{bm}
\long\def\begincomment#1\endcomment{}

\usepackage{xcolor} 

\begin{document}
\setlength{\baselineskip}{7mm}

\title { \textbf{
    Master stability functions for complete, intra-layer and inter-layer synchronization in multiplex networks}
}
\author{ Longkun Tang \footnotemark[1],
         Xiaoqun Wu \footnotemark[2] $^,$\footnotemark[4] $^,$ \footnotemark[5]
         Jinhu L\"u \footnotemark[3] ,
         Jun-an Lu \footnotemark[2] ,
         Raissa M. D'Souza \footnotemark[4] $^,$ \footnotemark[5]
         }
\renewcommand{\thefootnote}{\fnsymbol{footnote}}
 \footnotetext[1]{School of Mathematical Science, Huaqiao University, Quanzhou 362021, China. tomlk@hqu.edu.cn}
 \footnotetext[2]{School of Mathematics and Statistics, Wuhan University, Wuhan 430072, China.}
 \footnotetext[3]{Institute of Systems Science, Academy of Mathematics and Systems Science, Chinese Academy of Sciences, Beijing 100190, China.}
 \footnotetext[4]{Department of Computer Science, University of California, Davis CA 95616, USA.}
  \footnotetext[5]{To whom correspondence should be addressed: xqwu@whu.edu.cn, raissa@cse.ucdavis.edu }

\date{}
\maketitle{}

\begin{abstract}
Synchronization phenomena are of broad interest across disciplines and increasingly of interest in a multiplex network setting.  Here we show how the Master Stability Function, a celebrated framework for analyzing synchronization on a single network, can be extended to certain classes of multiplex networks with different intra-layer and inter-layer coupling functions. We derive three master stability equations that determine respectively the necessary regions of complete synchronization, intra-layer synchronization and inter-layer synchronization. We calculate these three regions explicitly for the case of a two-layer network of R{\"o}ssler oscillators and show that the overlap of the regions determines the type of synchronization achieved.
In particular, if the inter- or intra-layer coupling function is such that the inter-layer or intra-layer synchronization region is empty, complete synchronization cannot be achieved regardless of the coupling strength. Furthermore, for any given nodal dynamics and network structure, the occurrence of intra-layer and inter-layer synchronization depend mainly on the coupling functions of nodes within a layer and across layers, respectively. Our mathematical analysis requires that the intra- and inter-layer supra-Laplacians commute. But we show this is only a sufficient, and not necessary, condition and that the results can be applied more generally.
\end{abstract}


\section*{Introduction}

Synchronization in a network of connected elements is essential to the
proper functioning of a wide variety of natural and
engineered systems, from brain networks to electric power grids.
 This has stimulated a large number of investigations  into synchronization properties of complex networks, with small-world, scale-free and other types of topologies~\cite{WS1998,Barahona2002,Hong2002,Newman2003,Lu2004,Lu2005,Zhou2006,Wu2007,Boccaletti2006,Arenas2008,ChenY2011,Huang2008,
Donetti2005,Pecora2014, TangLK2012a}.
Yet many synchronization phenomena, as in electrical power grids, do not involve a single network in isolation but rely on the complete synchronization of a collection of smaller networks. And more generally, beyond single networks, we are now understanding that interactions between networks are increasingly important and that interactions can impact the dynamical processes~\cite{PNAS2012,Radicchi2013,Gomez2013,Domenico2014,Gregorio2014,Valles-Catala2016}. One paradigm that captures many real-world interdependent networks is that of multiplex networks. Here the same set of nodes exist in multiple layers of networks, where each layer represents a different interaction type,
the internal state of the corresponding nodes in each layer can be distinct, and the connectivity pattern between nodes in each layer can be distinct~\cite{kivela2014,Boccaletti2014}. As an example consider the online social system of a set of individuals. They may interact on Twitter or on Facebook or on Linked-in or on some combination of all three, and each layer can have its own  connectivity pattern, yet there is typically influence propagated between them. Given the need to study dynamical processes on layered complex networks, and the broad applicability of synchronization, here we study synchronization phenomena on multiplex networks,
an area that has attracted increasing attention in the past few years \cite{Ribalta2013,Aguirre2014,Boccaletti2014,Demomenico2015, LuRQ2014,Luo2014,Liyang2015,Xumm2015,Gambuzza2015,Sevilla2016}.

One of  the  most important methods to study network synchronization on single 
networks is the master stability function (MSF) method proposed by Pecora  and Carroll \cite{PC1998}. As established via the MSF approach, 
whether or not a network can achieve synchronization   is determined not only by the network structure,
but also by the nodal dynamics and
the inner coupling function which describes the
interactions among the different components of  the state vectors of connected nodes \cite{TangLK2012b,TangLK2014,TangLK2015}.
Current studies of synchronization phenomena in multiplex networks analyze a multiplex network as a single large composite network with the topology being described by a supra-Laplacian matrix. 
This requires that the inner coupling function is the same regardless of whether the nodes are linked by an intra-layer or inter-layer edge and the MSF framework can thus be directly applied.
The eigenvalues of this supra-Laplacian are then used to analyze the stability of the state of complete synchronization in multiplex networks.
For example, Sol\'{e}-Ribalta et al. \cite{Ribalta2013}  investigated the spectral properties of the Laplacian   of multiplex networks,
and discussed the synchronizability via the eigenratio of the Laplacian matrix.
Aguirre et al. \cite{Aguirre2014} studied the impact of the connector node degree on the synchronizability of two star networks with one inter-layer link
and showed that connecting the high-degree (low-degree) nodes of each network is the most (least) effective strategy to achieve synchronization.
Xu et al. \cite{Xumm2015} investigated the synchronizability of two-layer networks for three specific coupling patterns,
and determined that there exists an optimal value of the inter-layer coupling strength for maximizing complete synchronization in the two-layer networks they analyze.
Li et al. \cite{Liyang2015} investigated the sychronizability of a duplex network composed of two star networks with two inter-layer links
by giving an analytical expression containing the largest and the smallest nonzero eigenvalues of the Laplacian matrix, the link weight, as well as the network size.

In 2012, Sorrentino et al.~\cite{sorrentinoNJP, sorrentinoPRE} considered an innovative  ``hypernetwork" model
consisting of one set of $N$ nodes that interact via multiple types of coupling functions.
Note the contrast with a multiplex network, where a set of $N$ nodes exists on each one of $M$ distinct layers (for a total of $M \times N$ nodes), and each node can be in a different state in each layer. (See for instance, Fig. \ref{inter_intra}.)
In the ``hypernetwork" model there are only $N$ nodes in total and each node can be in only one state at any given time. As such, the focus is on complete synchronization and
three situations are found where the network topology is such that one can
decouple the effects of interaction functions from the structure of the networks 
and apply the MSF approach~\cite{sorrentinoNJP, sorrentinoPRE}.
Extremely recently, del Genio et al. extended this analysis to a broader range of scenarios, again using an MSF approach~\cite{Genio2016}, and show how the ``hypernetwork" model of \cite{sorrentinoNJP, sorrentinoPRE} is equivalent to a network where nodes have many different interaction types (or ``layers" of interaction).  Although these works consider that
nodes can interact with one another via different coupling functions, they do not capture the richness of phenomena that can occur in multiplex networks
such as intra-layer and inter-layer synchronization.

Only limited studies thus far have focused on intra-layer and inter-layer synchronization.
For example,
Gambuzza et. al~\cite{Gambuzza2015} analyzed synchronization of a population of oscillators indirectly coupled through   an inhomogeneous medium.  The system is formalised in terms of a two-layer network, where the top layer is composed of disconnected oscillators, and the bottom layer consists of oscillators  coupled according to a given topology, and each node in the top layer is connected to its counterpart in the bottom layer.  By numerical simulations, they have shown the onset of intra-layer synchronization  without inter-layer coherence, that is,  a state in which the nodes of a layer are synchronized between them without being synchronized with those of the other layer.  Shortly afterwards,  Sevilla-Escoboza et. al \cite{Sevilla2016} investigated the inter-layer synchronization in a duplex network of identical layers, and 
showed that there are instances
where each node in a given layer can synchronize with  its replica in the other layer irrespective of whether or not intra-layer synchronization occurs.
These findings into specific systems provide useful foundations for elucidating
a more fundamental approach to analyzing synchronization phenomena in multiplex networks. In fact, as we will show herein, master stability equations can be derived to systematically predict when intra- and inter-layer synchronization are simultaneously supported and when they are not simultaneously supported for certain classes of multiplex networks.

Based on the above motivations, here we develop a Master Stability Function method which captures an essential feature of multiplex networks, that the inter-layer coupling function can be distinct from the intra-layer coupling function.
Thus, distinct from previous approaches, we can analyze different kinds of coherent behaviors,
 including   complete synchronization,  intra-layer synchronization and inter-layer  synchronization in multiplex networks, however, we are restricted to certain classes of topologies.
{In particular, we derive  the master stability equation for  a multiplex network where the supra-Laplacian of intra-layer connections and that of inter-layer connections commute, as defined in detail below. We further derive
two reduced forms of the master stability equation corresponding to only inter-layer or intra-layer interactions. We then show how three different necessary regions for synchronization can be calculated from the MSF of the three master stability equations.  Finally we show how to explicitly apply the multiplex MSF by analyzing
a specific example of two-layer network of R\"ossler oscillators  with identical intra-layer topological structures and one-to-one inter-layer connections.
For broader applicability of this multiplex MSF approach, we further illustrate  that the three master stability equations can still be used to predict the area of synchronization for some classes of multiplex networks with non-commutative supra-Laplacians.

\section*{Results}\label{framework}
\noindent\textbf{A multiplex network model.}
We consider a multiplex network consisting of $M$ layers each consisting of  $N$ nodes.
The state of the $i$-th node  in the $k$-th layer is specified by $\bm{x}_i^{(k)}=(x_{i1}^{(k)},x_{i2}^{(k)},\cdots, x_{im}^{(k)})^\top $,
an $m-$dimensional state vector. The evolution of the full multiplex system  can be written as: 
\begin{equation}\label{MLNet0}
\ \dot {\bm{x}_i}^{(k)} = f(\bm{x}_i^{(k)})-c \sum\limits_{j = 1}^N
l_{ij}^{(k)}H({\bm{x}_j}^{(k)})-d\sum\limits_{l = 1}^M d_{kl}\Gamma (\bm{x}_i^{(l)}),
\quad i=1,2,\cdots,N;\,k=1,2,\cdots,M ,\
\end{equation}
where $\dot {\bm{x}_i}^{(k)} = f(\bm{x}_i^{(k)})\;(i=1,2,\cdots,N;\,k=1,2,\cdots,M)$
describes the isolated dynamics for the $i$-th node  in the $k$-th layer,
and $f(\cdot):\mathbb{R}^m\rightarrow \mathbb{R}^m$ is a well-defined vector function,
$H(\cdot):\mathbb{R}^m\rightarrow \mathbb{R}^m$  and $c$ are the  inner coupling function  and coupling strength  for nodes  within each layer, respectively,
and $\Gamma(\cdot):\mathbb{R}^m\rightarrow \mathbb{R}^m$  and $d$  are  the inner coupling function and coupling strength for nodes across  layers, respectively.
For simplicity and clarity, we let $H(\bm{x})=H\bm{x}$ and $\Gamma(\bm{x})=\Gamma\bm{x}$, namely we set the coupling functions between nodes to be linear (thus we can also call $H$ and $\Gamma $ inner coupling matrices).
Furthermore,  the inner coupling matrix for nodes within one layer, $H$, is identical for all layers and the inner coupling matrix for nodes across two layers,
$\Gamma$, is the same for all pairs of layers.

Elements $l_{ij}^{(k)}$ describe the Laplacian matrix of nodes within  the $k$-th layer. Explicitly, if the $i$-th node is connected with the $j$-th node within the $k$-th layer, $l_{ij}^{(k)}=-1$, otherwise $l_{ij}^{(k)}=0$,
and  $l_{ii}^{(k)}= - \sum_{j=1}^N l_{ij}^{(k)}$, for $ i, j=1,2,\cdots,N  $ and $k=1,2,\cdots,M. $ Similarly,
if a node in the $k$-th layer is connected with its replica in the $l$-th layer,  $d_{kl}=-1$,
otherwise $d_{kl}=0$, and  $d_{kk}= - \sum_{l=1}^M d_{kl}$, for $k,l=1,2,\cdots,M. $

For simplicity, denote \\
$\bm{x}^{(k)}=\left(\begin{array}{c}
  \bm{x}_{1}^{(k)} \\
  \bm{x}_{2}^{(k)} \\
  \vdots \\
  \bm{x}_{N}^{(k)}
\end{array}\right),$
$\tilde{f}(\bm{x}^{(k)})=
\left(\begin{array}{c}
  f(\bm{x}_{1}^{(k)}) \\
  f(\bm{x}_{2}^{(k)}) \\
  \vdots \\
  f(\bm{x}_{N}^{(k)})
\end{array}\right),$
$\mathbf{x}=\left(\begin{array}{c}
  \bm{x}^{(1)} \\
  \bm{x}^{(2)} \\
  \vdots \\
  \bm{x}^{(M)}
\end{array}\right),$
$F(\mathbf{x})=
\left(\begin{array}{c}
  \tilde{f}(\bm{x}^{(1)}) \\
  \tilde{f}(\bm{x}^{(2)}) \\
  \vdots \\
  \tilde{f}(\bm{x}^{(M)})
\end{array}\right),$ \\
then the evolution of the multiplex network (Eq.~\ref{MLNet0}) can be rewritten as
\begin{equation}\label{MLNet1}
\ \dot {\mathbf{x}} = F(\mathbf{x})-c(\mathcal{L}^L\otimes H)\mathbf{x}
    -d(\mathcal{L}^I\otimes\Gamma)\mathbf{x},\
\end{equation}
where $\mathcal{L}^L$ stands for the supra-Laplacian of intra-layer connections and $\mathcal{L}^I$ for the  supra-Laplacian of inter-layer connections. In detail,
$\mathcal{L}^L=\bigoplus_{l=1}^M {L}^{(k)}
       =\left(\begin{array}{cccc}
               L^{(1)} &  &  &  \\
                & L^{(2)} &  &  \\
                &  & \ddots &  \\
                &  &  & L^{(M)}
               \end{array}
\right)$ and
$\mathcal{L}^I=L^I\otimes I_N$. Here $\bigoplus$ is the direct sum operation, $I_N$ is the N-by-N identity matrix,
$\otimes$ is the Kronecker product operation,
$L^{(k)}=(l_{ij}^{(k)})_{N\times N}$ is the Laplacian matrix of nodes within  the $k$-th layer,
and $L^I=(d_{kl})_{M\times M}$ represents  the inter-layer Laplacian matrix.
More details about supra-Laplacians and multiplex network models can be found in Refs. \cite{Gomez2013,Ribalta2013,Boccaletti2014,Liyang2015,Xumm2015}
and  references therein.

\vspace{1cm}
\noindent\textbf{A master stability function framework for classes of multiplex networks.}
The master stability function method~\cite{PC1998} is one of the most important methods to study stability of synchronized coupled identical systems.
It simplifies a large-scale networked system to a node-size system
via diagonalization and decoupling, as long as the inner coupling functions for all node pairs are identical.
Thus, determining   whether a network can reach  synchronization
can be turned into determining  whether all the network characteristic modes fall into the corresponding synchronized regions.
In the following, we will establish a master stability framework for multiplex networks with nonidentical inter-layer and intra-layer inner coupling functions.

According to the idea of the master stability framework \cite{PC1998}, to investigate network synchronization,
we can linearize the dynamical equation (\ref{MLNet1}) at $\bm{1}_M\otimes\bm{1}_N\otimes\bm{s}$,
where $\bm{s}$ is a synchronous  state of the network satisfying $\dot{\bm{s}}=f(\bm{s})$ and $\bm{1}_M $ denotes  an $M$-dimensional vector with all entries being 1.
We thus  obtain the following variational equation:
\begin{equation}\label{VarEq0}
  \dot{\bm{\xi}}=[I_{M\times N}\otimes Df(\bm{s})-c(\mathcal{L}^L\otimes H)
    -d(\mathcal{L}^I\otimes\Gamma)]\bm{\xi},
\end{equation}
where $\bm{\xi}=\mathbf{x}-\bm{1}_M\otimes\bm{1}_N\otimes\bm{s}$
and $I_{M\times N}$ is the identity matrix of order $M\times N$.

Suppose that $\mathcal{L}^L$ and $\mathcal{L}^I$ are symmetric matrices, and satisfy
$\mathcal{L}^L\mathcal{L}^I=\mathcal{L}^I\mathcal{L}^L$.  
{After diagonalization and decoupling} {(see the Methods section for details)}, we get  the multiplex master stability equation for  a system described by Eq. (\ref{MLNet0}):
\begin{equation}\label{VarEq3}
  \dot{\bm{y}}=[Df(\bm{s})-\alpha H
    -\beta\Gamma]\bm{y},
\end{equation}
where $\alpha=c\lambda$, $\beta=d\mu$,
$\lambda$ and $\mu$ are the eigenvalues of $\mathcal{L}^L$ and $\mathcal{L}^I$ respectively,
and satisfy $\lambda^2 + \mu^2 \neq 0$.

Since this equation may be a time-varying system, particularly if $s(t)$ is a function of time, its eigenvalues may not be useful for determining the stability. Therefore, the largest Lyapunov exponent (LLE) of Eq. (\ref{VarEq3}) is used instead, which is a function of $\alpha$ and $\beta$, denoted $LLE(\alpha, \beta)$ and called the multiplex Master Stability Function for Eq. (\ref{MLNet0}). 

When $\lambda\neq 0$ and $\mu=0$, there is no inter-layer couplings regardless of $d$,
for $d$ arbitrarily chosen in $[0, \, +\infty)$, and Eq. (\ref{VarEq3}) reduces to
\begin{equation}\label{VarEq4}
  \dot{\bm{y}}=[Df(\bm{s})-\alpha H]\bm{y},
\end{equation}
It is clear that Eq. (\ref{VarEq4}) becomes exactly the  master stability equation
of each independent intra-layer network  (no inter-layer couplings).

Similarly, when $\lambda=0$ and $\mu\neq0$,  we can obtain the following equation
\begin{equation}\label{VarEq5}
  \dot{\bm{y}}=[Df(\bm{s})-\beta\Gamma]\bm{y},
\end{equation}
regardless of  coupling strength $c$,  for $c$ arbitrarily chosen in $[0, \, +\infty)$.
Eq. (\ref{VarEq5}) becomes exactly the master stability equation for each independent inter-layer network (no intra-layer couplings).

For  a single layer network,  a necessary condition for the synchronization manifold to be stable is that the largest Lyapunov exponent $LLE(\alpha)$ of Eq. (\ref{VarEq4}) less than zero \cite{PerronEff2007}. In analogy to a single layer,  for the multiplex master stability equation (\ref{VarEq3}),  $LLE(\alpha, \beta) <0$ is a necessary condition for  stability of the   synchronization manifold  in a multiplex network.

It is worth noting in particular the case when the intra- and inter-layer coupling functions are identical. Here $H=\Gamma$, and Eq. (\ref{VarEq3}) turns into $\dot{\bm{y}}=[Df(\bm{s})-\gamma H]\bm{y}$,
(with  $\gamma=\alpha+\beta$).
This is the master stability equation for the corresponding single composite network
where a single supra-Laplacian can describe its topology.
That is to say, the master stability equation of the single composite network is a special case of Eq.  (\ref{VarEq3}).


 The assumption that  $\mathcal{L}^L$ and $\mathcal{L}^I$ are symmetric
and satisfy the commutativity condition is an important condition
for decoupling the system and restricts our approach from applying to the full class of multiplex networks.
But this assumption can be relaxed, as it is only a sufficient but not a necessary condition.
First we consider the case when $\mathcal{L}^L$ and $\mathcal{L}^I$ are commutative but are nonsymmetric.  As shown in the Supplementary Information, the same master stability equations (\ref{VarEq3})-(\ref{VarEq5})
(which correspond to Eqs.~(S10), (S9) and (S6), respectively)
can be derived
provided that the multiplex network has intra-layer topology that is identical on each layer and that both the intra-layer Laplacian matrix $L^L$ and the inter-layer Laplacian matrix $L^I$ can be diagonalizable and have real eigenvalues.
There are important classes of real-world networks that fit this paradigm,
such as the CORS geospatial information infrastructure~\cite{Rizos2007,Blewitt1998,LiuH2016} discussed in detail in the Supplementary Information.

More generally, we also study the case when $\mathcal{L}^L$ and $\mathcal{L}^I$ do not commute. As shown in detail later in the simulation results section, for a two-layer Rossler network with non-commutative supra-Laplacians
the three master stabilty equations (\ref{VarEq3})-(\ref{VarEq5}) can be still used to predict network synchronization behaviors.
In particular, for duplex networks, if the network topology is different on each layer, but there is one-to-one identical weighted coupling of nodes between layers, we can predict complete synchronization and intra-layer synchronization. If the topology on each layer is identical, but the one-to-one weighted coupling is not identical, we can predict complete synchronization and inter-layer synchronization.

\vspace{1cm}
\noindent\textbf{Synchronized regions.}
Using the multiplex master stability equations developed above,
we can analyze three types of synchronization behaviors: complete synchronization, intra-layer synchronization and inter-layer synchronization.  Here we define the regions that support each behavior and in the
subsequent sections and Supplemental Information we show that it is the overlap of these regions that determines the type of sychronization pattern displayed by a multiplex network.

For the full multiplex network, from the multiplex master stability equation (\ref{VarEq3})
we can calculate the region
\begin{equation*}
 SR=\{(\alpha,\beta)|LLE(\alpha,\beta)<0\},
\end{equation*}
which is called the joint synchronized region (which supports complete synchronization of the network). Whenever $LLE(\alpha,\beta)<0$,
perturbations transverse to the 
synchronization manifold die out, and the network is said to be synchronizable.

From Eq. (\ref{VarEq4}), we can get the region for intra-layer synchronization.
The region depends only on the value of the parameter $\alpha$, but to later allow comparison across the full parameter space we explicitly include the parameter $\beta$ in the definition of the region,
\begin{equation*}
  SR^{\,Intra}_{\alpha, \beta}=\{(\alpha,\beta)|\,LLE(\alpha)<0 \},
\end{equation*}
where $LLE(\alpha)$ is the largest Lyapunov exponent for master stability equation (\ref{VarEq4}).
Similarly, from Eq. (\ref{VarEq5}), we obtain the region for inter-layer synchronization 
\begin{equation*}
SR^{\,Inter}_{\alpha, \beta}=\{(\alpha,\beta)|LLE(\beta)<0 \}.
\end{equation*}
We call these regions in the parameter space of $(\alpha \geq 0,\beta \geq 0)$ the corresponding synchronized regions with respect to $\alpha$ and $\beta$.

When the network topological structures are specified, we can determine $\lambda$ and $\mu$ (the eigenvalues of $\mathcal{L}^L$ and $\mathcal{L}^I$) directly, and then the regions $SR$, $SR^{\,Intra}_{\alpha,\beta}$ and $SR^{\,Inter}_{\alpha,\beta}$  can be parameterized 
simply in terms of coupling strengths $c$ and $d$, denoted by $SR_{c, d}$, $SR_{c,d}^{\,Intra}$ and $SR_{c,d}^{\,Inter}$.
We call these regions the corresponding synchronized regions with respect to couplings $c$ and $d$. 

\vspace{1cm}
\noindent\textbf{Two-layer network of R\"{o}ssler oscillators.}
With the multiplex MSF framework developed, we now analyze in more depth a specific example of a two-layer network
of R\"{o}ssler oscillators and calculate the different types of synchronized regions.

The famous R\"{o}ssler chaotic oscillator is described as
\begin{equation}\label{rossler}
\left\{\begin{array}{l}
       \dot x = -y - z, \\
       \dot y = x + ay,  \\
       \dot z = z (x - c) + b \,,  \\
\end{array}\right.
\end{equation}
where $a=b=0.2$ and $c=9$. This is the function $f$ in the multiplex network Eq. (\ref{MLNet0}). That is, the state of each node in the network is a three-dimensional vector with each component evolving by Eq. (\ref{rossler}).
For inner coupling matrices $H$ and $\Gamma$, we consider the family of choices that fit the simplest form   $I_{ij}\in \mathbb{R}^{3\times 3}~({\rm where\ } i,j=1,2,3)$,  which represents  a matrix whose $(i,j)$-element is one and other elements are zero. The inter-layer topology is set to be one-to-one connection, that is to say, each node in one layer  is connected to a counterpart node in the other layer.

\vspace{1cm}
\noindent\textbf{Intra-layer and inter-layer synchronization.}
It is well known that complete synchronization means all the nodes in a network come to an identical state.
But for multiplex networks, it is also very significant to study intra-layer synchronization and inter-layer synchronization.
As shown in Fig. \ref{inter_intra}, intra-layer synchronization means all the nodes within each layer reach an identical state, while inter-layer synchronization means each node in a layer reaches the same state as its counterparts in other layers.

\begin{figure}[!ht]
  \centering
 \includegraphics[width=16cm]{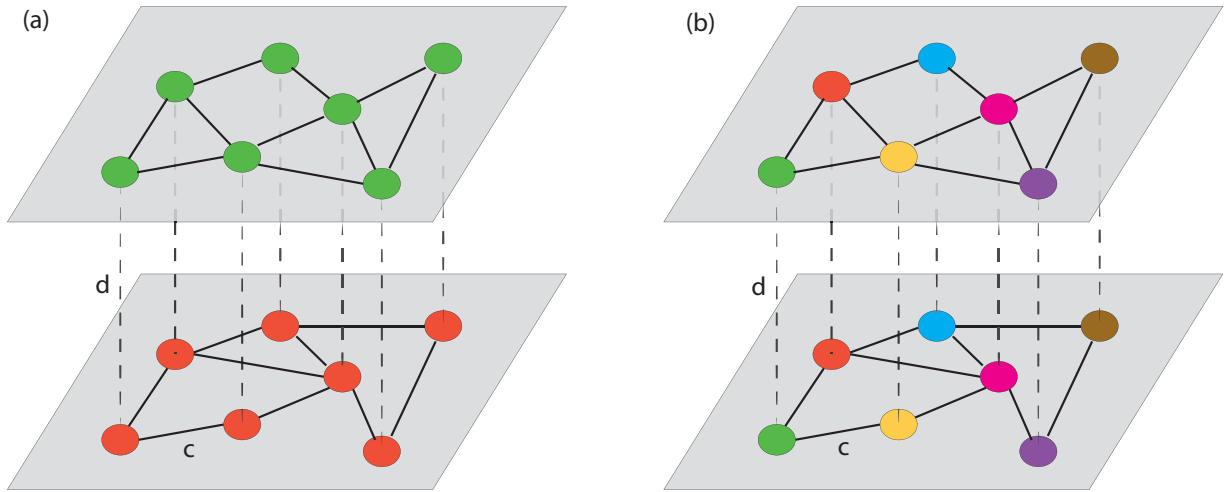}\\
     \caption{Schematic representation of  (a) intra-layer synchronization and (b) inter-layer synchronization, in a multiplex network of two layers.}\label{inter_intra}
\end{figure}

\vspace{1cm}
\noindent\textbf{Synchronized regions for unknown intra-layer topologies.}
The regions of synchronization calculated from the multiplex MSF are parameterized by $\alpha$ and $\beta$, and thus do not require that the inter- and intra-layer topology are specified.
Figure~\ref{Regions} shows the
synchronized regions as parameterized by ($\alpha, \beta$)
for a  two-layer multiplex network of R\"{o}ssler oscillators with arbitrary topology 
for different combinations of inter-layer and intra-layer coupling matrices $H$ and $\Gamma$.
Here, the green shading represents the regions $SR$ as obtained from the master stability equation (\ref{VarEq3}).
The regions $SR^{\,Intra}_{\alpha,\beta}$ as obtained from Eq. (\ref{VarEq4}) are enclosed by the blue dash-dotted lines,
and the regions $SR^{\,Inter}_{\alpha,\beta}$ as obtained from Eq. (\ref{VarEq5}) enclosed by the red dash-dotted lines.
\begin{figure}[!ht]
  \centering
 \includegraphics[width=16cm]{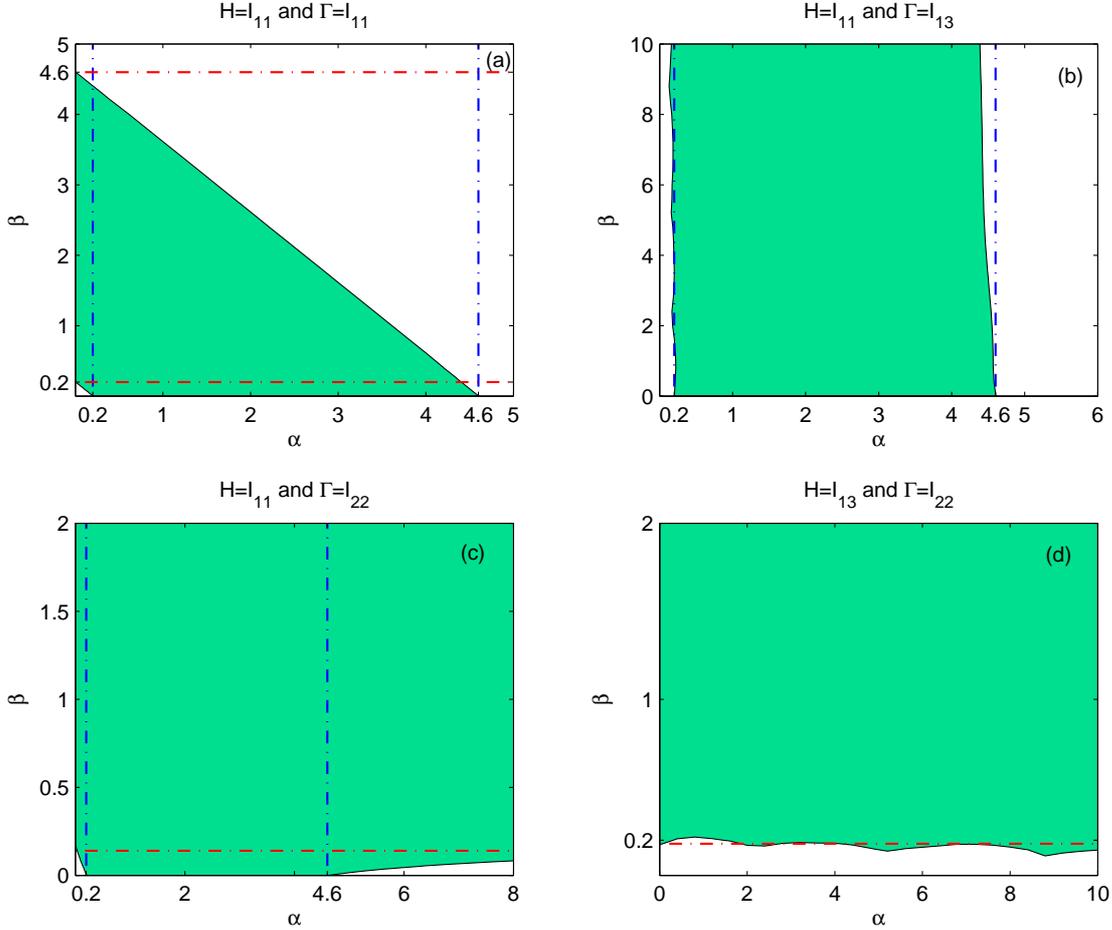}\\ 
    \caption{  The synchronized regions with respect to $\alpha$ and $\beta$, $SR$ painted with green color,
    $SR^{\,Intra}_{\alpha,\beta}$ enclosed by the blue dash-dotted lines, and $SR^{\,Inter}_{\alpha,\beta}$ enclosed by the red dash-dotted lines.
  Here the R\"{o}ssler oscillator is taken as nodal dynamics, and the intra-layer coupling matrix $H$
  and the inter-layer coupling matrix $\Gamma$ are chosen as follows:
  (a) $H=I_{11}$, $\Gamma=I_{11}$, (b) $H=I_{11}$, $\Gamma=I_{13}$,
  (c) $H=I_{11}$, $\Gamma=I_{22}$, (d) $H=I_{13}$, $\Gamma=I_{22}$.
   }\label{Regions}
\end{figure}

Synchronization occurs in the region when the MSF criterion  is negative,  in other words when $LLE(\alpha, \beta) < 0$.
Thus from Fig.~\ref{Regions}, we can easily obtain the joint synchronized region
$SR=\{(\alpha,\beta)|\,0.2<\alpha+\beta<4.6\}$ for $H=I_{11}$ and $\Gamma=I_{11}$,
$SR\approx\{(\alpha,\beta)|\,0.23<\alpha<4.3,\beta\geq0\}$ for $H=I_{11}$ and $\Gamma=I_{13}$,
$SR\approx\{(\alpha,\beta)|\,\frac{\alpha}{0.2}+\frac{\beta}{0.18}>1, \beta>-10^{-5}\alpha^4
   +0.00057\alpha^3-0.012\alpha^2+0.12\alpha-0.35\}$ for $H=I_{11}$ and $\Gamma=I_{22}$,
and $SR\approx\{(\alpha,\beta)|\,\beta>0.2, \alpha\geq0\}$ for $H=I_{13}$ and $\Gamma=I_{22}$.

In particular,   letting $\beta=0$ in $SR$, we have $\{\alpha|\,   LLE(\alpha)<0\}=(0.2, \,4.6)$ for $H=I_{11}$,
 $\{\alpha|\, LLE(\alpha)<0\}=\emptyset$ for $H=I_{13}$, and $\{\alpha|\, LLE(\alpha)<0\}=(0.18,\,\infty)$ for $H=I_{22}$.
Similarly,  letting $\alpha=0$ in $SR$, we have $\{\beta|\, LLE(\beta)<0\}=(0.2, \,4.6)$ for $\Gamma=I_{11}$,
$\{\beta|\, LLE(\beta)<0\}=\emptyset$ for $\Gamma=I_{13}$, and $\{\beta|\, LLE(\beta)<0\}=(0.18,\,\infty)$ for $\Gamma=I_{22}$.
Here, the intervals $\{\alpha|\, LLE(\alpha)<0\}$ and $\{\beta|\, LLE(\beta)<0\}$ can   also be obtained
from Eqs. (\ref{VarEq4}) and (\ref{VarEq5}), respectively.

Consequently, $SR^{\,Intra}_{\alpha,\, \beta}=\{(\alpha,\beta)|\,0.2<\alpha<4.6,\,\beta\geq 0\}$ for $H=I_{11}$ and any $\Gamma$,
$SR^{\,Intra}_{\alpha,\, \beta}=\emptyset$ for $H=I_{13}$ and any $\Gamma$,
and $SR^{\,Intra}_{\alpha,\, \beta}=\{(\alpha,\beta)|\,0.18<\alpha<+\infty,\,\beta\geq 0\}$ for $H=I_{22}$ and any $\Gamma$.
Analogously, we can obtain $SR^{\,Inter}_{\alpha,\, \beta}$  by replacing   $\alpha$ with  $\beta$, and $H$ with $\Gamma$ in the above $SR^{\,Intra}_{\alpha,\, \beta}$.

As shown in Fig.~\ref{Regions}, for $H=I_{11}$ and $\Gamma=I_{11}$,
the regions $SR^{\,Intra}_{\alpha,\, \beta}=\{(\alpha,\beta)|\,0.2<\alpha<4.6,\,\beta\geq 0\}$
and $SR^{\,Inter}_{\alpha,\, \beta}=\{(\alpha,\beta)|\,0.2<\beta<4.6,\,\alpha\geq 0\}$,
which are enclosed by the blue and red dash-dotted lines, respectively. For $H=I_{11}$ and $\Gamma=I_{13}$,
$SR^{\,Intra}_{\alpha,\, \beta}=\{(\alpha,\beta)|\,0.2<\alpha<4.6,\,\beta\geq 0\}$, enclosed by the blue dash-dotted lines,
and $SR^{\,Inter}_{\alpha,\, \beta}=\emptyset$. For $H=I_{11}$ and $\Gamma=I_{22}$,
$SR^{\,Intra}_{\alpha,\, \beta}=\{(\alpha,\beta)|\,0.2<\alpha<4.6,\,\beta\geq 0\}$, enclosed by the blue dash-dotted lines,
and $SR^{\,Inter}_{\alpha,\, \beta}=\{(\alpha,\beta)|\,0.18<\beta<+\infty,\,\alpha\geq 0\}$, which is above the red dash-dotted line.
For $H=I_{13}$ and $\Gamma=I_{22}$, $SR^{\,Intra}_{\alpha,\, \beta}=\emptyset$,
and $SR^{\,Inter}_{\alpha,\, \beta}=\{(\alpha,\beta)|\,0.18<\beta<+\infty,\,\alpha\geq 0\}$, which is above the red dash-dotted line.

Generally speaking, a multiplex network with a specified topology can achieve complete synchronization when
all the nonzero network characteristic modes, including those of the intra-layer and inter-layer Laplacians,
fall into the synchronized region.
For a two-layer network with identical intra-layer topologies,
our theoretical analysis (see the Supplementary Information) further shows that a duplex network can achieve
complete synchronization when all the nonzero characteristic modes fall into the intersection of  $SR$,   $SR^{\,Intra}_{\alpha,\, \beta}$ and $SR^{\,Inter}_{\alpha,\, \beta}$.  Therefore, according to the overlapping region, one can determine  whether the network   achieves complete synchronization
or not after specifying the topology. However, what happens when  all the nonzero characteristic modes do not fall into the intersection?
Further simulations shows that in this case the network could support other coherent dynamical behaviors, such as intra-layer or inter-layer synchronization.

\vspace{1cm}
\noindent\textbf{Synchronized regions  with given intra-layer topologies.}
 To push the analysis further, we must specify the topology of the
 two-layer R\"{o}ssler oscillator network. For simplicity,
assume that the two layers have the same intra-layer topology,
and each node in one layer is connected with its replica in the other layer.
Consider that each layer is a star network consisting of 5 nodes. Then, the intra-layer Laplacian matrix is
\begin{equation*}
  L=\left(
      \begin{array}{ccccc}
        4 & -1 & -1 & -1 & -1 \\
        -1 & 1 & 0 & 0 & 0 \\
        -1 & 0 & 1 & 0 & 0 \\
        -1 & 0 & 0 & 1 & 0 \\
        -1 & 0 & 0 & 0 & 1 \\
      \end{array}
    \right),
\end{equation*}
and the intra-layer supra-Laplacian matrix is
$\mathcal{L}^L=\left(
                 \begin{array}{cc}
                   L & 0 \\
                   0 & L \\
                 \end{array}
               \right)$.
The inter-layer Laplacian matrix
$L^I=\left(
    \begin{array}{cc}
          1 & -1 \\
          -1 & 1 \\
          \end{array}
          \right)$,
and the inter-layer supra-Laplacian matrix $\mathcal{L}^I=L^I\otimes I_5$.
It is easy to verify that $\mathcal{L}^L\mathcal{L}^I=\mathcal{L}^I\mathcal{L}^L$,
and the characteristic values of $\mathcal{L}^L$ and $\mathcal{L}^I$ are $\lambda=0,\,0,\,1,\,1,\,1,\,1,\,1,\,1,\,5,\,5$
and $\mu=0,\,0,\,0,\,0,\,0,\,2,\,2,\,2,\,2,\,2$, respectively.

We can calculate the eigenvalues $\lambda$ and $\mu$ directly and parameterize the synchronized regimes by the coupling strengths $c$ and $d$ (rather than the more general $\alpha$ and $\beta$) for all the different combinations of the inner coupling matrices $H$ and $\Gamma$.
(See the Methods section for more details on transforming $SR$ to $SR_{c.d}$.)
Consequently, for $H=I_{11}$ and $\Gamma=I_{11}$, the region with respect to parameters  $c$ and $d$ is
$$SR_{\,c,d}=\{(c, d)|\,0.2<c+2d, c+0.4d<0.92\}.$$
Similarly, for $H=I_{11}$ and $\Gamma=I_{13}$, then $SR_{\,c,d}\approx\{(c,d)|\,0.23<c<0.86, d\geq0\}$;
for $H=I_{11}$ and $\Gamma=I_{13}$, then $SR_{\,c,d}\approx\{(c,d)|\,\frac{c}{0.2}+\frac{d}{0.09}>1, d>\frac{1}{2}(-625\cdot10^{-5}c^4
   +0.07125c^3-0.2c^2+0.6c-0.35)\}$;
and for $H=I_{13}$ and $\Gamma=I_{22}$, then $SR_{\,c,d}\approx\{(c,d)|\,d>0.1, c\geq0\}$.
These regions, $SR_{c,d}$, are shown  in  Figs.~\ref{RosH11T11all}-\ref{RosH13T22all}
by the solid lines in panels (c) for the different choices of $H$ and $\Gamma$ considered.
\begin{figure}[!ht]
  \centering
  \includegraphics[width=16cm]{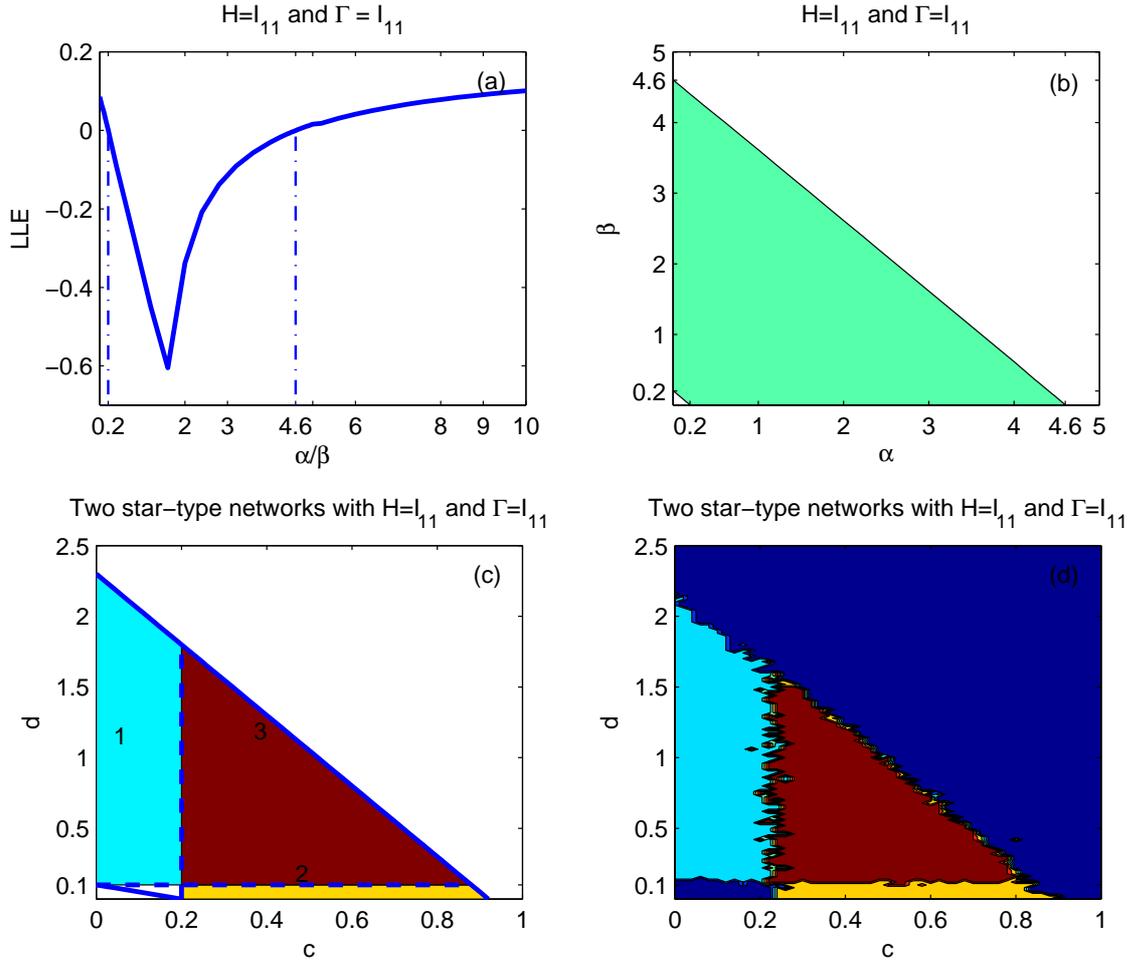}\\
  \caption{ Network synchronized regions for $H=I_{11}$ and $\Gamma=I_{11}$.
   (a) the synchronized interval of the independent intra-layer/inter-layer R\"{o}ssler network with respect to $\alpha$/$\beta$;
  (b) the synchronized region with respect to $\alpha$ and $\beta$ for  R\"{o}ssler networks;
  (c) the synchronized region with respect to couplings $c$ and $d$ for a R\"{o}ssler duplex
  consisting of two star layers with one-to-one inter-layer connections;
  (d) numerical  synchronization areas  with respect to couplings $c$ and $d$,
  in which the maroon region represents complete synchronization area, the yellow is for intra-layer synchronization,
  and the cyan is inter-layer synchronization and the blue region represents non-synchronization.}\label{RosH11T11all} 
\end{figure}

\begin{figure}[!ht] 
  \centering
  \includegraphics[width=16cm]{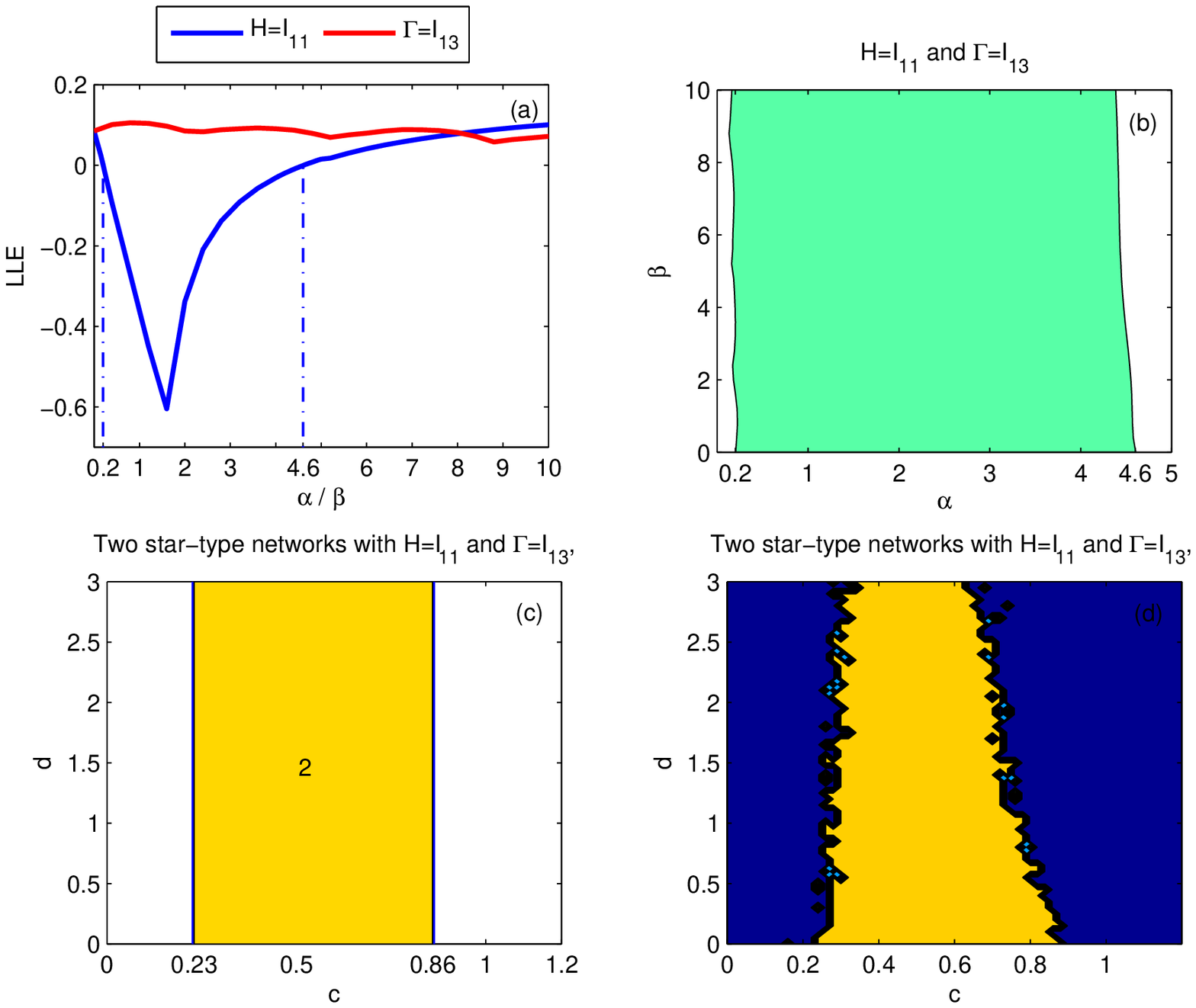}\\
  \caption{Network synchronized regions for $H=I_{11}$ and $\Gamma=I_{13}$.
  (a) the synchronized interval of the independent intra-layer/inter-layer R\"{o}ssler network with respect to $\alpha$/$\beta$;
  (b) the synchronized region with respect to $\alpha$ and $\beta$ for R\"{o}ssler networks;
  (c) the synchronized region with respect to couplings $c$ and $d$ for a  R\"{o}ssler duplex  consisting of two  star layers with one-to-one inter-layer connections;
  (d)  numerical  synchronization areas   with respect to couplings $c$ and $d$, in which the maroon region represents complete synchronization area,
  the yellow is for intra-layer synchronization, and
  the cyan is inter-layer synchronization and the blue region represents non-synchronization.
  }\label{RosH11T13all}
\end{figure}

To test our theoretical predictions we next numerically solve the duplex R\"{o}ssler networked system, and identify the parameter regions that support the three different coherent behaviors: complete synchronization, intra-layer synchronization and inter-layer synchronization.  We quantify that the system has reached the specific type of behavior via the
synchronization errors as defined in the Methods section.
By bounding the values of these errors we develop three different indicator functions, which identify that the system has achieved macroscopic order of the form:
Id=3 when the network reaches complete synchronization, Id=2  for  intra-layer synchronization,  Id=1 for  inter-layer synchronization and Id=0  for none of the above cases. See Methods for full details.  

Figure~\ref{RosH11T11all} shows network synchronized regions for the two-layer star network of R\"{o}ssler oscillators for the scenario  $H=I_{11}$ and $\Gamma=I_{11}$.
In detail, panel (a) displays the synchronized intervals of the independent intra-layer and inter-layer R\"{o}ssler networks with respect to $\alpha$ or $\beta$, which can be calculated from the master stability equations (\ref{VarEq4}) and (\ref{VarEq5}) (without consideration of $d$ or $c$), respectively.  Since $H = \Gamma$, the two intervals overlap.
Panel (b) gives the synchronized region with respect to $\alpha$ and $\beta$
for  this R\"{o}ssler network calculated from the master stability equation (\ref{VarEq3}).
Panel (c) shows the synchronized region as a function of intra- and inter-layer coupling strength $c$ and $d$.
Panel (d) shows  the numerically calculated indicator function} (i.e., the numerically calculated  values of synchronization error as classified in Eq.~(\ref{indicator}) given in the Methods section) with respect to couplings $c$ and $d$ for this duplex R\"{o}ssler network. Here the maroon area represents complete synchronization, the yellow area is for intra-layer synchronization, the cyan is for inter-layer synchronization, and the blue region represents cases otherwise.

\begin{figure}[!ht] 
  \centering
  \includegraphics[width=16cm]{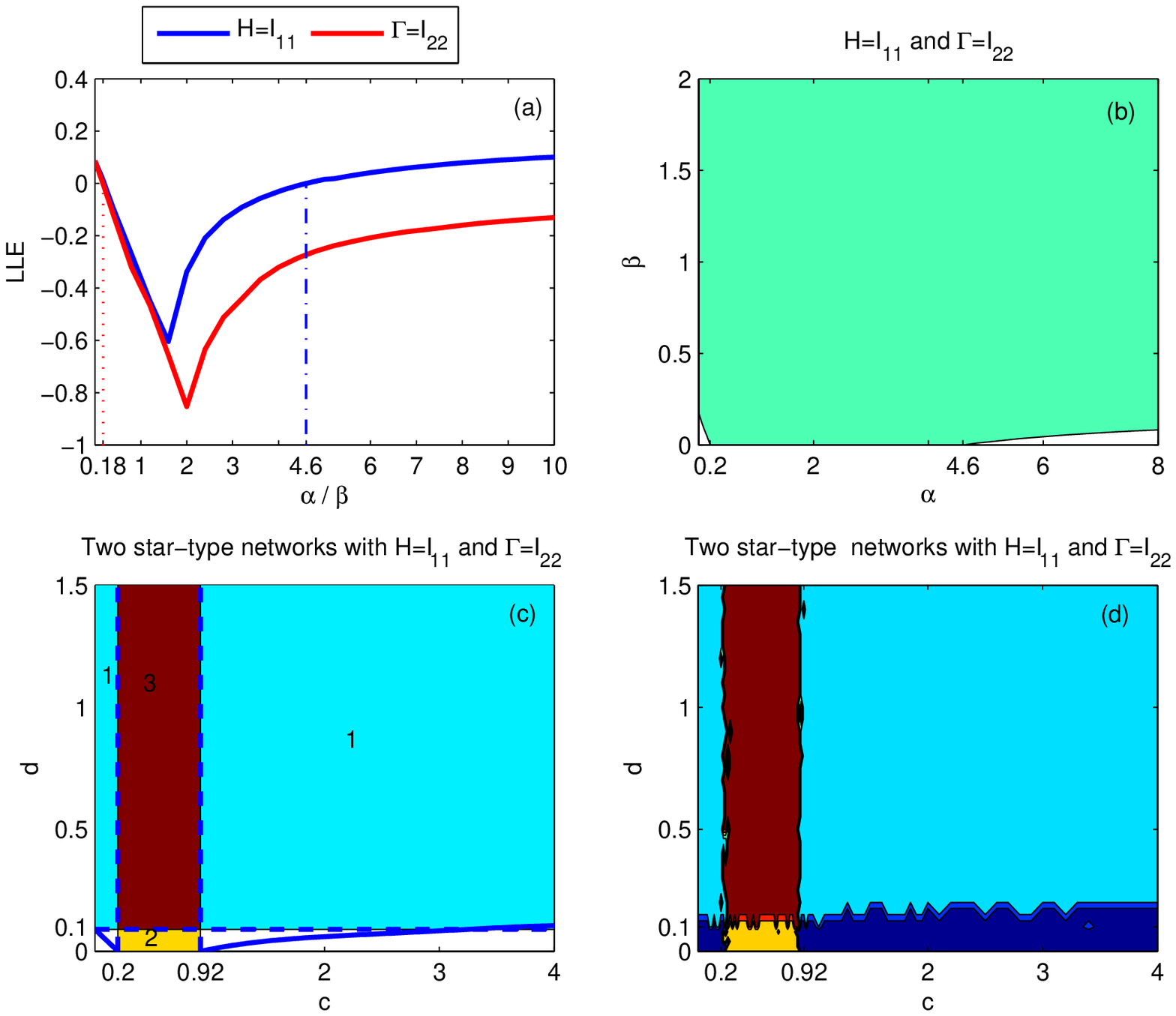}\\
  \caption{Network  synchronized regions for $H=I_{11}$ and $\Gamma=I_{22}$.
  (a) the synchronized interval of the independent intra-layer/inter-layer R\"{o}ssler network with respect to $\alpha$/$\beta$;
  (b) the synchronized region with respect to $\alpha$ and $\beta$ for  R\"{o}ssler networks;
  (c) the synchronized region with respect to couplings $c$ and $d$
  for a   R\"{o}ssler duplex  consisting of two  star layers with one-to-one inter-layer connections;
  (d)  numerical  synchronization areas   with respect to couplings $c$ and $d$,
  in which the maroon region represents complete synchronization area,
  the yellow is for intra-layer synchronization, and
  the cyan is inter-layer synchronization and the blue region represents non-synchronization.}\label{RosH11T22all}
\end{figure}

\begin{figure}[!ht] 
  \centering
  \includegraphics[width=16cm]{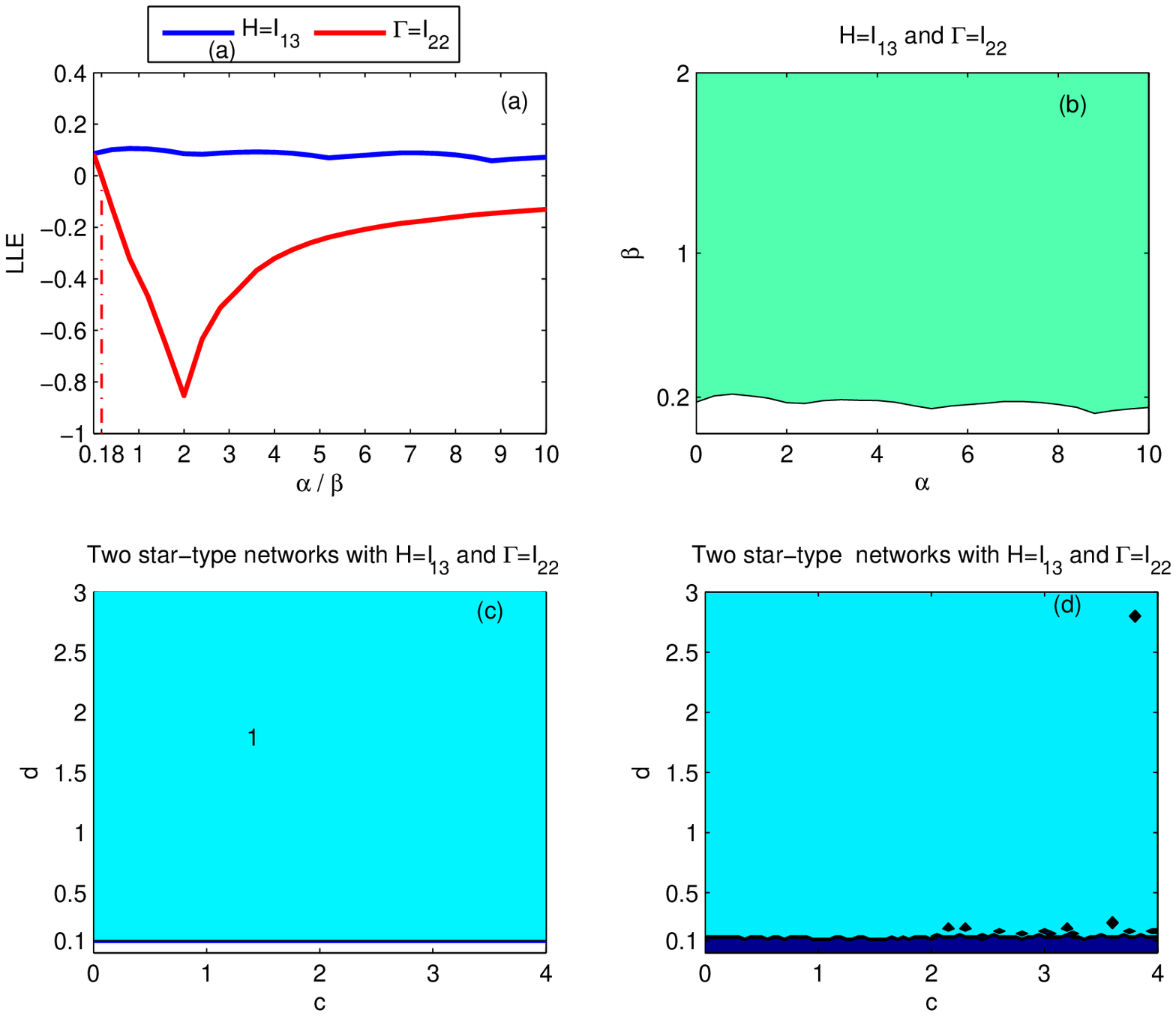}\\
  \caption{Network  synchronized regions for $H=I_{13}$ and $\Gamma=I_{22}$.
  (a) the synchronized interval of the independent intra-layer/inter-layer R\"{o}ssler network with respect to $\alpha$/$\beta$;
  (b) the synchronized region with respect to $\alpha$ and $\beta$ for  R\"{o}ssler networks;
  (c) the synchronized region with respect to couplings $c$ and $d$
  for a   R\"{o}ssler duplex  consisting of two  star layers with one-to-one inter-layer connections;
  (d)  numerical  synchronization areas   with respect to couplings $c$ and $d$,
  in which the maroon region represents complete synchronization area,
  the yellow is for intra-layer synchronization,
  and the cyan is inter-layer synchronization and the blue region represents non-synchronization.}\label{RosH13T22all}
\end{figure}

\begin{figure}[!ht] 
  \centering
  \includegraphics[width=16cm]{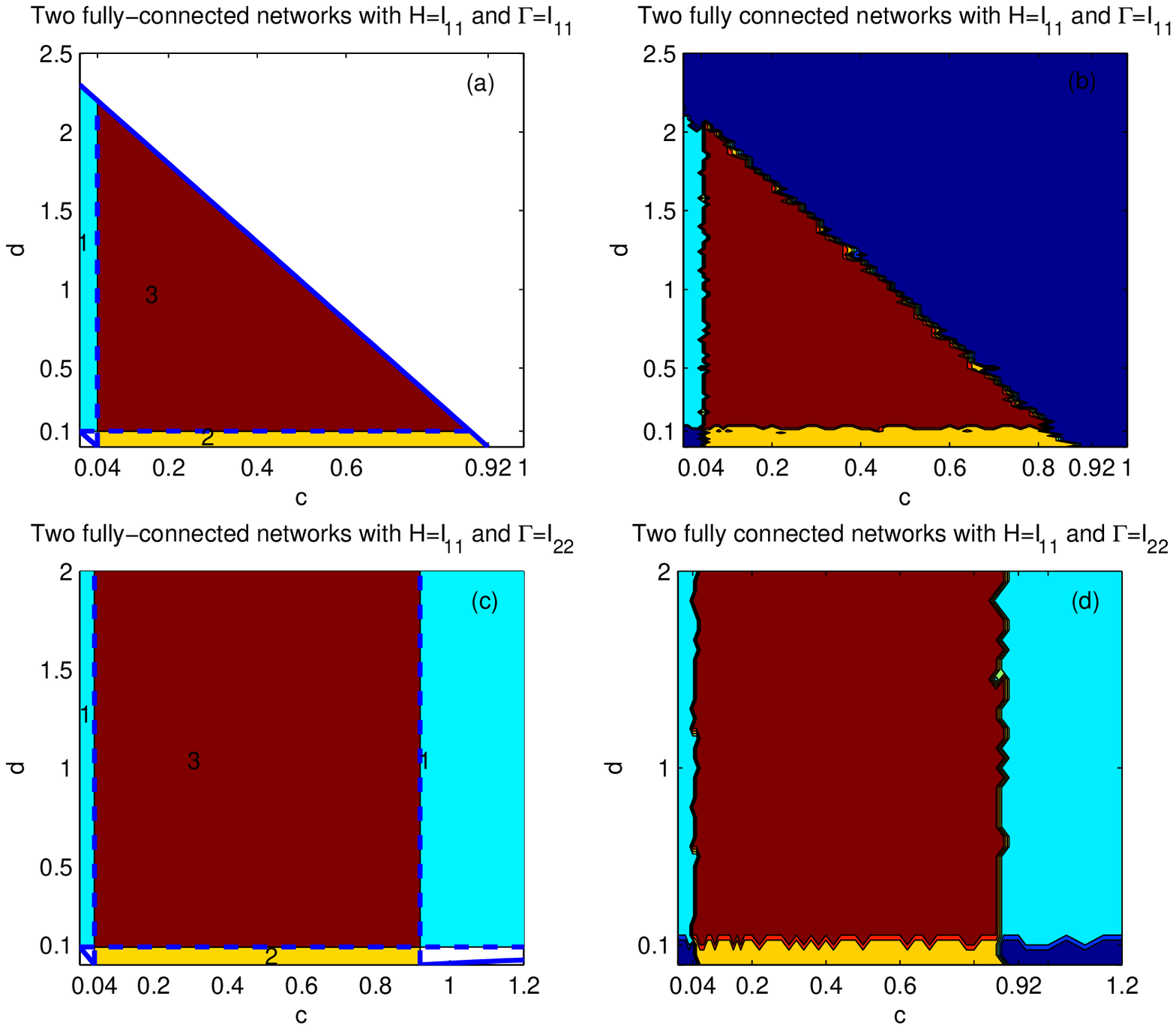}\\
  \caption{Network  synchronized regions for R\"{o}ssler networks composed of
  two single-layer fully connected networks with different $H$ and $\Gamma$.
  (a) and (c) are the synchronized regions about $c$ and $d$; (b) and (d) are numerical synchronization areas,
  in which the maroon region means complete synchronization, the yellow region means mere intra-layer synchronization,
  the cyan region means mere inter-layer synchronization and the blue region means non-synchronization.}\label{RosH11T11H11T22all}
\end{figure}

\begin{figure}[!ht] 
  \centering
  \includegraphics[width=16cm]{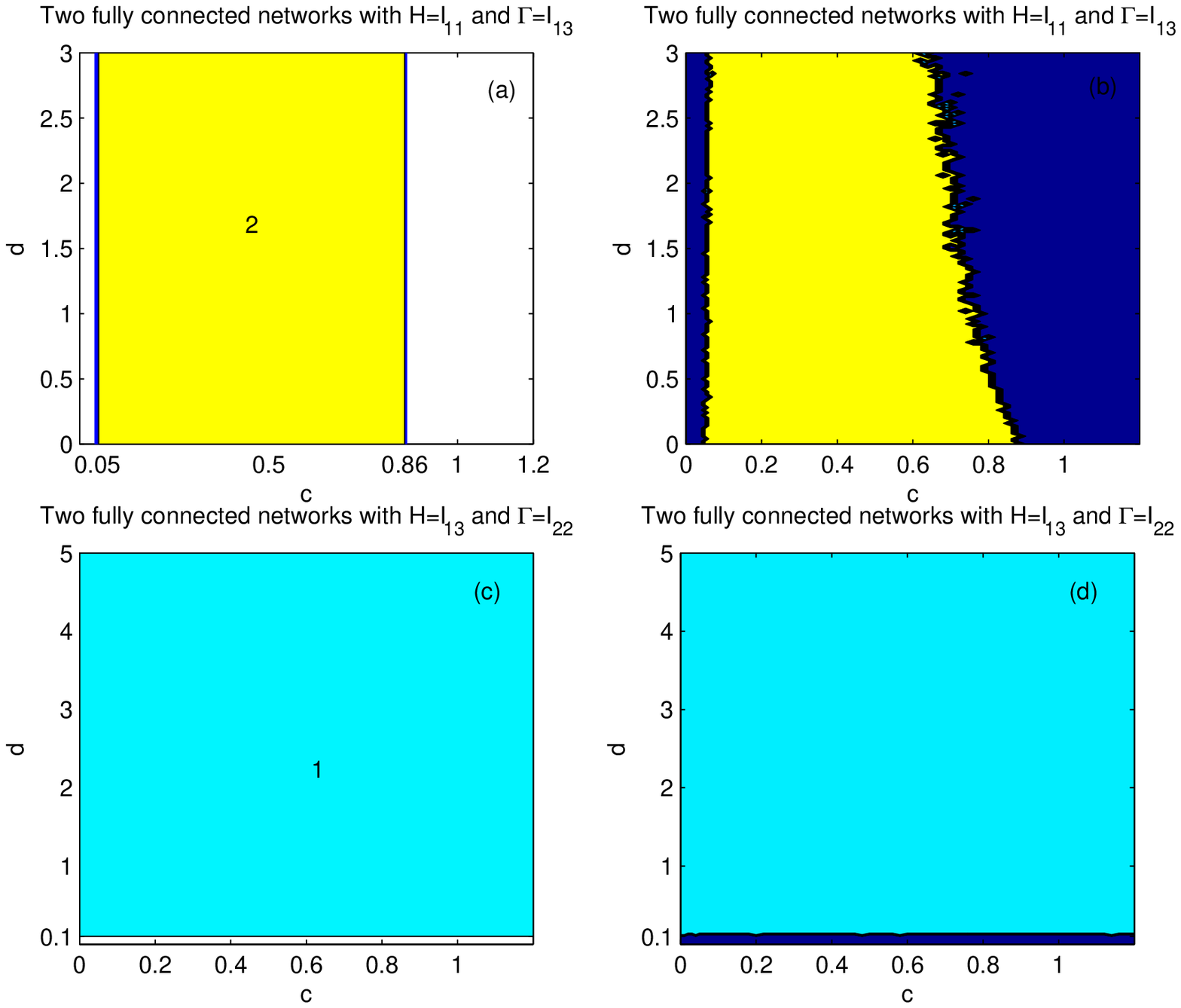}\\
  \caption{Network  synchronized regions for R\"{o}ssler networks composed of
  two single-layer fully connected networks with different $H$ and $\Gamma$.
  (a) and (c) are the synchronized regions about $c$ and $d$; (b) and (d) are numerical synchronization areas,
  in which the maroon region means mere complete synchronization, the yellow region means mere intra-layer synchronization,
  the cyan region means inter-layer synchronization and the blue region means non-synchronization.
  }\label{RosH11T13H13T22all}
\end{figure}

Other choices for the coupling functions $H$ and $\Gamma$ are shown in Figures \ref{RosH11T13all}-\ref{RosH13T22all}
for this same two-layer star network of R\"{o}ssler oscillators. The results are analogous to those in Fig. \ref{RosH11T11all}.
It is worth noting  that in panels (c) of  all of these figures the regions of complete, intra-layer and inter-layer synchronization predicted by the multiplex MSF Eqs. (\ref{VarEq3}) to  (\ref{VarEq5}), shown as the maroon, yellow and cyan regions respectively, can  capture all of the behaviors exhibited by direct numerical simulations shown in panels (d).

Next we show how these distinct areas can be determined from the three regions: $SR_{c,d}$,  $SR_{c,d}^{Intra}$, and $SR_{c,d}^{Inter}$  derived from Eqs.~(\ref{VarEq3}), (\ref{VarEq4}), and (\ref{VarEq5}).
As a matter of fact, the intersections of the regions determine the type of coherent behavior that is stable.
Specifically, the intersection of all the three regions determines complete synchronization,
the intersection of $SR_{c,d}$ and $SR_{c,d}^{Intra}$ determines intra-layer synchronization,
and the intersection of $SR_{c,d}$ and $SR_{c,d}^{Inter}$ determines inter-layer synchronization.

For example, for the case with $H=I_{11}$ and $\Gamma=I_{11}$,  the synchronized region $SR_{c,d}=\{(c, d)|\, c+2d>0.2, c+0.4d<0.92\}$, the intra-layer synchronized region
$SR_{c,d}^{Intra}=\{(c, d)|\,0.2<c<0.92,\,d\geq 0\}$ and the inter-layer synchronized region $SR_{c,d}^{Inter}=\{(c, d)| c\geq 0,  \,0.1<d<2.3\, \}$.
The  intersection of these three parts is $\{(c, d)|\,c>0.2,\; d>0.1,\; c+0.4d<0.92\}$, as labeled by number `3' in  panel (c) of Fig.~\ref{RosH11T11all},
which essentially 
coincides with the numerically calculated complete synchronization area in maroon color in panel (d).
Furthermore,  the mere intra-layer synchronization (without inter-layer synchronization) area in yellow  in panel (d)
coincides with the region labeled as `2' in panel (c):
$SR_{c,d}\cap SR_{c,d}^{Intra}-SR_{c,d}^{Inter}=\{(c, d)|\,c>0.2, 0 \leq d<0.1,\, c+0.4d<0.92\}$,
and the  mere inter-layer synchronization (without intra-layer synchronization) area in cyan agrees well with the region labeled as `1' in panel (c):
$SR_{c,d}\cap SR_{c,d}^{Inter}-SR_{c,d}^{Intra}=\{(c, d)|\,0 \leq c<0.2,\, d>0.1,\, c+0.4d<0.92\}$.
Similar observations can be obtained in panels (c) and (d) of Figs. \ref{RosH11T13all}-\ref{RosH13T22all}.

In other words, the actual area for complete synchronization is determined by the intersection of
$SR_{c,d}$, $SR_{c,d}^{Intra}$ and $SR_{c,d}^{Inter}$, that is, $SR_{c,d}\cap SR_{c,d}^{Intra}\cap SR_{c,d}^{Inter}$.
Moreover, the mere intra-layer synchronization  area is determined by the intersection of  synchronized region and intra-layer snchronized region subtracting the inter-layer synchronized part, that is,  $SR_{c,d}\cap SR_{c,d}^{Intra}-SR_{c,d}^{Inter}$.
The mere inter-layer synchronization area is determined by $SR_{c,d}\cap SR_{c,d}^{Inter}-SR_{c,d}^{Intra}$.

Furthermore, when nodal dynamics and network structures are given,
$SR_{c,d}^{Intra}$ and $SR_{c,d}^{Inter}$ are mainly determined by the inner coupling matrices of the intra-layer nodes ($H$) and the inter-layer nodes ($\Gamma$) respectively, and $SR_{c,d}$ is determined by both.
Particularly, if the inter-layer coupling matrix $\Gamma$ makes the inter-layer synchronized region $SR_{c,d}^{Inter}$ empty,
then the multiplex network cannot achieve inter-layer synchronization, resulting in  the failure of complete synchronization,
 as shown in Fig. \ref{RosH11T13all}.
If the intra-layer coupling matrix $H$ makes the intra-layer synchronized region $SR_{c,d}^{Intra}$ empty,
then the multiplex network cannot achieve intra-layer synchronization, which also leads to  failure of complete synchronization, as shown in Fig.  \ref{RosH13T22all}.

In order to verify the previous results on a different multiplex topology,
we   consider a duplex network composed of two fully connected network layers with one-to-one inter-layer connections.
The results  shown in Figs. \ref{RosH11T11H11T22all} and \ref{RosH11T13H13T22all} again illustrate the above  observations.

\begin{figure}[!ht] 
  \centering
  \includegraphics[width=16cm]{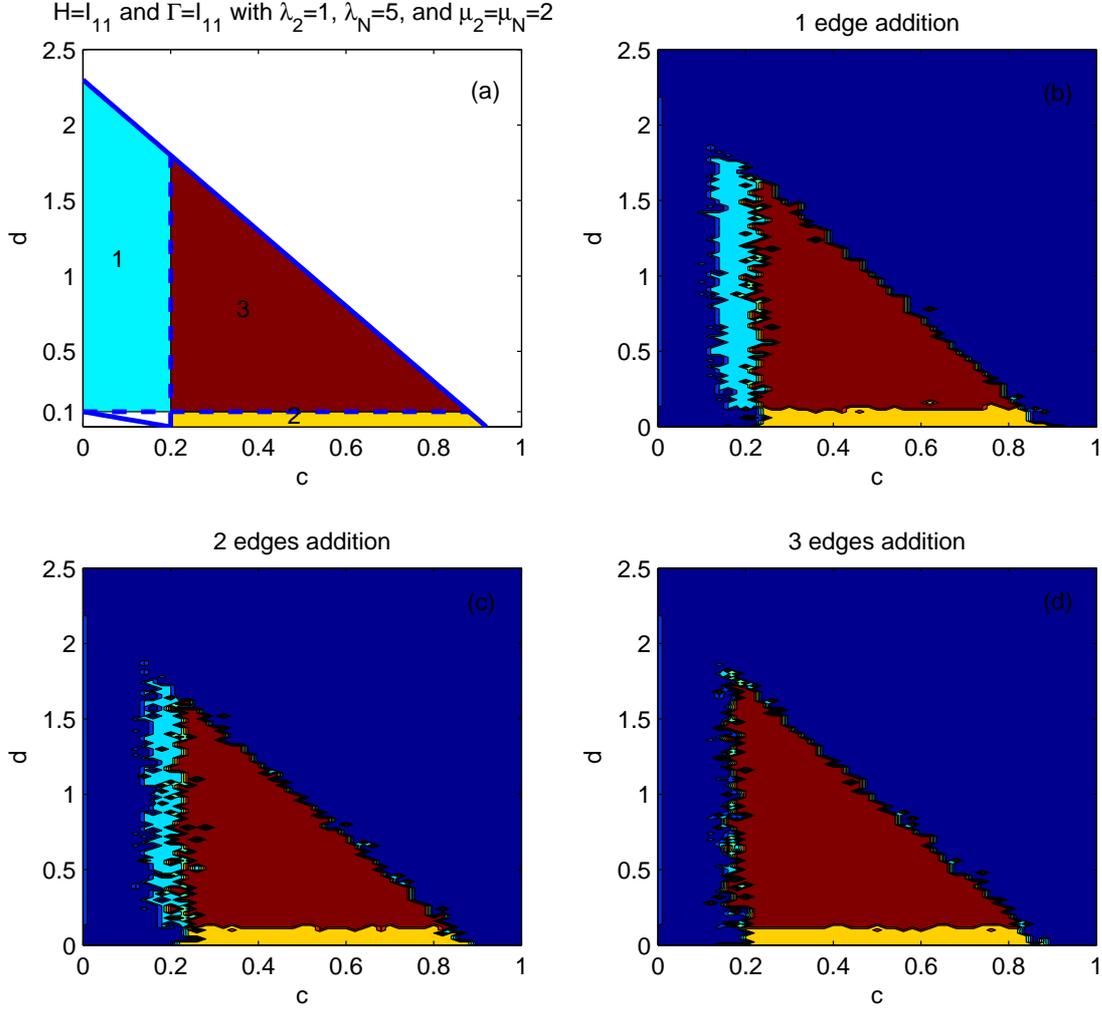}\\
  \caption{\footnotesize The case of non-commutative supra-Laplacian matrices with different intra-layer topologies and identical inter-layer coupling weights.
   Network synchronized regions calculated from master stability equations (a)
   and  the  numerical synchronization areas (b)(c)(d) for $H=I_{11}$ and $\Gamma=I_{11}$.
   The first layer of the duplex network is the star-type, the second layer is the one generated from the star-type
   with 1 (b), 2 (c) and 3 (d) additional edges, respectively. The one-to-one coupling between layers is identical.
  }\label{RosH11T11csDL}
\end{figure}

\begin{figure}[!ht] 
  \centering
  \includegraphics[width=16cm]{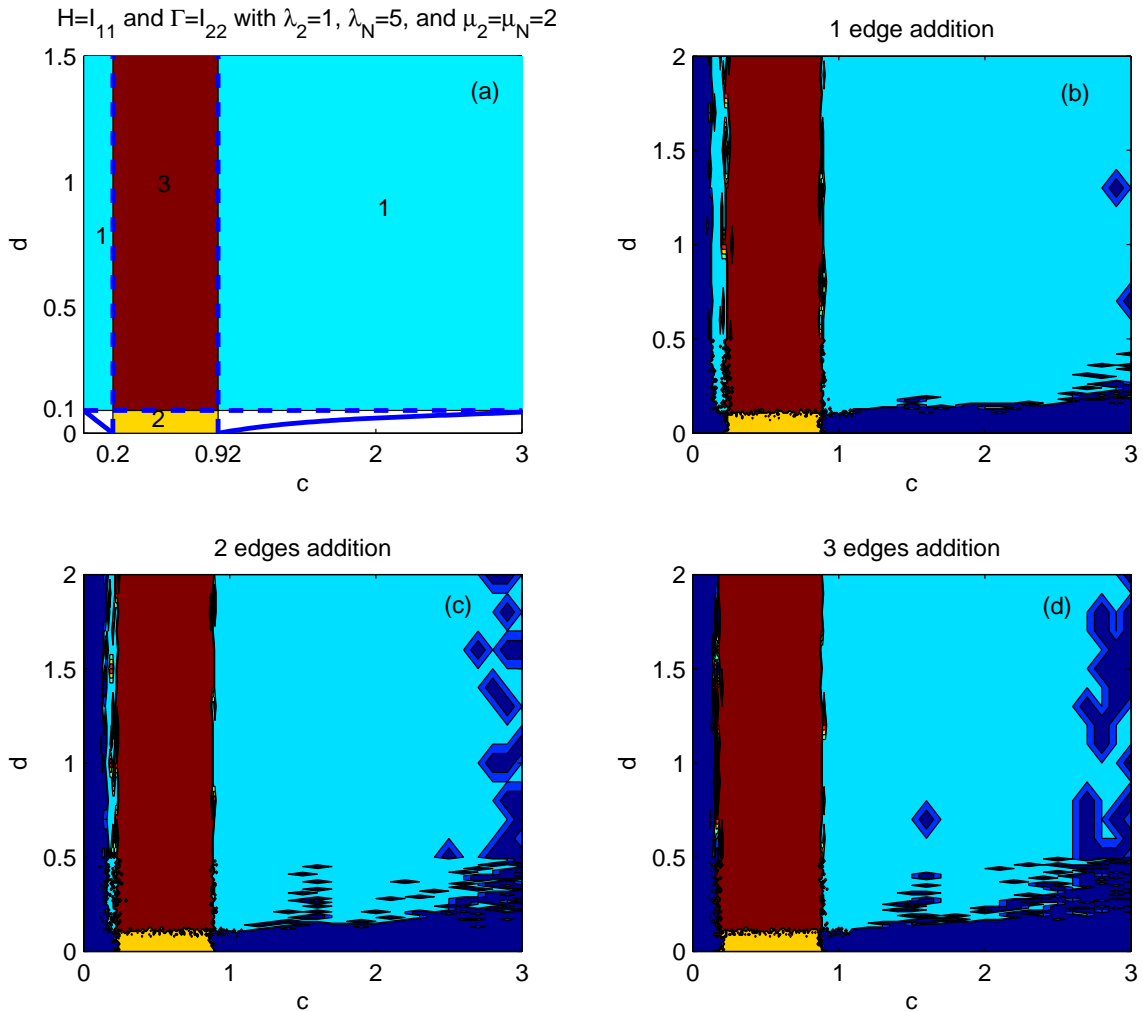}\\
  \caption{\footnotesize The case of non-commutative supra-Laplacian matrices with different intra-layer topologies and identical inter-layer coupling weights.
   Network synchronized regions calculated from master stability equations (a)
   and  the  numerical synchronization areas (b)(c)(d) for  $H=I_{11}$ and $\Gamma=I_{22}$.
   The first layer of the duplex network is the star-type, the second layer is the one generated from the star-type
   with 1 (b), 2 (c) and 3 (d) additional edges, respectively. The one-to-one coupling between layers is identical.
  }\label{RosH11T22csDL}
\end{figure}

\vspace{1cm}
\noindent\textbf{Two-layer R\"{o}ssler network with non-commutative supra-Laplacians.}
So far we have analyzed the case of commutative supra-Laplacians with which we derive the three master stability equations (\ref{VarEq3})-(\ref{VarEq5}).
However, the commutativity condition is only a sufficient but not a necessary condition, and can be relaxed.

Here we consider two cases of non-commutative supra-Laplacians to show that the three master stability equations can be still used to predict the area of synchronization for some classes of multiplex networks.
One case is a duplex network that has different topology on each layer and one-to-one identical weighted coupling of nodes between layers.
The other case is a duplex network that has identical topology on each layer and one-to-one nonidentical weighted coupling of nodes between layers.

For the first case, consider a specific duplex networks with 5 nodes on each layer and one-to-one coupling of nodes between layers, where one layer is the star-type,
and the other is the star-type with 1, 2 or 3 additional edges.
In this case, it is easy to verify that the intra- and inter-layer supra-Laplacian matrices $\mathcal{L}^L=\left( \begin{array}{cc}
              L_1& 0 \\
              0 & L_2
            \end{array}\right)$
and $\mathcal{L}^I=L^I\otimes I_N$
do not commute. Here
$L^I=\left( \begin{array}{cc}
              1& -1 \\
              -1 & 1
            \end{array}\right)$,
and the smallest nonzero eigenvalues of the two intra-layer Lapacian matrices $L_1$ and $L_2$ are equal and
their largest eigenvalues are also equal, i.e., $\lambda_2=1$ and $\lambda_N=5$.

For the second case, consider specific duplex networks that have identical star-type or fully-connected topology on each layer,
and nonidentical weighted one-to-one coupling between layers, with the inter-layer supra-Lapalacian matrix being
$\mathcal{L}^I=L^I\otimes\mbox{diag}\{2,1,1,1,1\}$, where here $L^I=\left( \begin{array}{cc}
              1& -1 \\
              -1 & 1
            \end{array}\right)$.
In this case, the smallest nonzero eigenvalue of $\mathcal{L}^I$ is $\mu_2=2$ and its largest eigenvalue is $\mu_N=4$.
It is easy to verify that $\mathcal{L}^L$ and $\mathcal{L}^I$ do not commute.

\begin{figure}[!ht] 
  \centering
  \includegraphics[width=13cm]{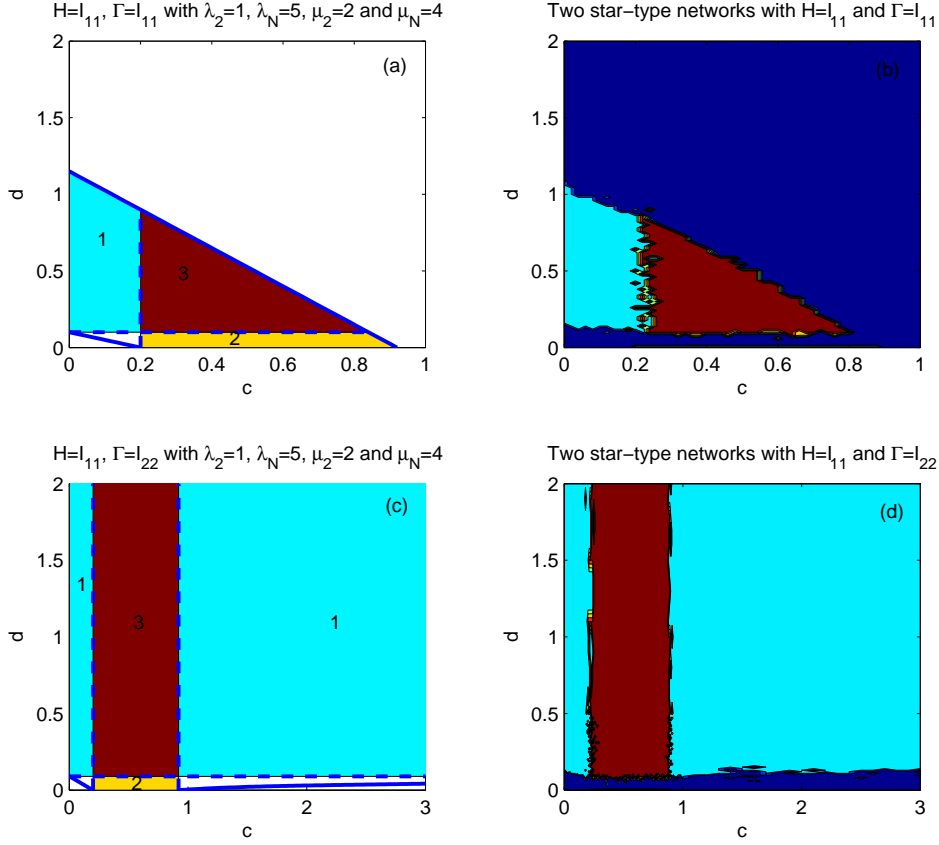}\\
  \caption{\footnotesize The case of non-commutative supra-Laplacian matrices. Network synchronized regions from master stability equations (left) and numerical synchronization areas (right) for R\"{o}ssler networks with identical star-type intra-layer topologies and nonidentical one-to-one coupling weights
  between layers. Here the inter-layer supra-Laplacian matrix $\mathcal{L}^I=[1 \; -1; -1\;1]\otimes\mbox{diag}\{2,1,1,1,1\}$,
  $H=I_{11}$ and $\Gamma=I_{11}$ for (a)(b), and $H=I_{11}$ and $\Gamma=I_{22}$ for (c)(d).
  }\label{S_RosH11T11H11T22cs_I2}
\end{figure}

\begin{figure}[!ht] 
  \centering
  \includegraphics[width=13cm]{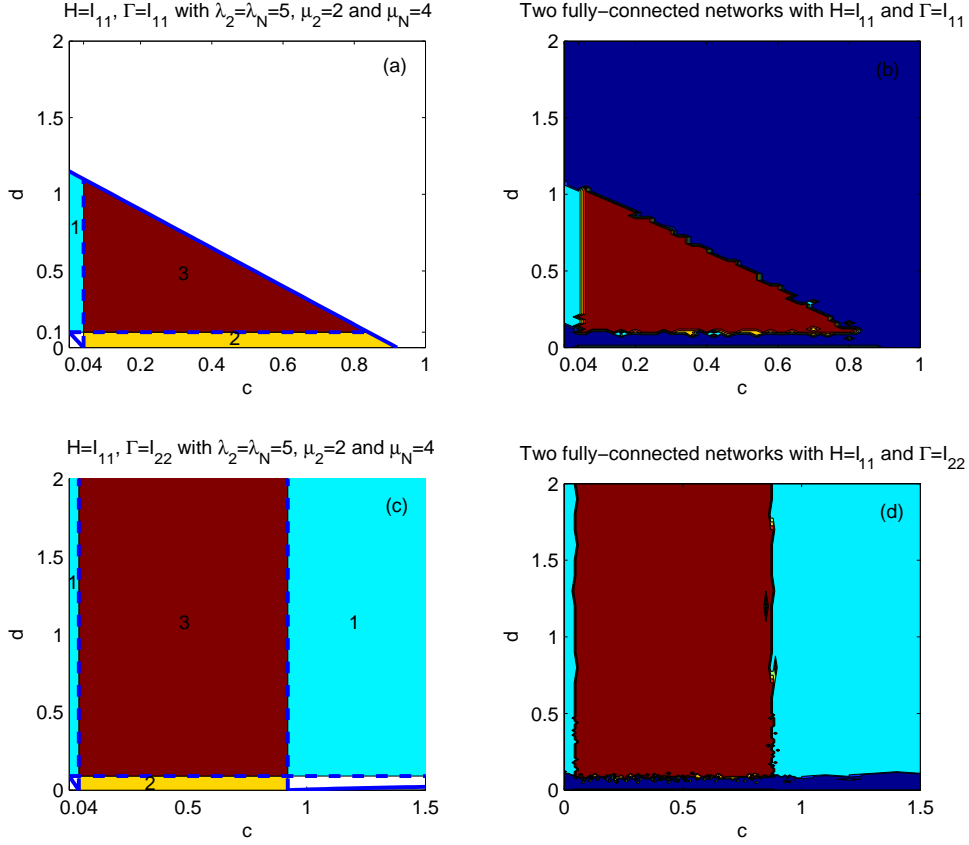}\\
  \caption{\footnotesize The case of non-commutative supra-Laplacian matrices. Network synchronized regions from master stability equations (left) and numerical synchronization areas (right) for R\"{o}ssler networks with identical fully-connected intra-layer topologies and nonidentical one-to-one coupling weights between layers. Here the inter-layer supra-Laplacian matrix $\mathcal{L}^I=[1 \; -1; -1\;1]\otimes\mbox{diag}\{2,1,1,1,1\}$,
  $H=I_{11}$ and $\Gamma=I_{11}$ for (a)(b), and $H=I_{11}$ and $\Gamma=I_{22}$ for (c)(d).
  }\label{F_RosH11T11H11T22cs_I2}
\end{figure}

Figures \ref{RosH11T11csDL}--\ref{F_RosH11T11H11T22cs_I2} show results for the above two different classes of duplex networks with different combinations of $H$ and $\Gamma$.
We still find that the overlapping regions obtained from the three master stability equations closely
coincide with the numerically calculated areas for the three different types of synchronous behaviors.
Specifically, the actual area for complete synchronization is determined by $SR_{c,d}\cap SR_{c,d}^{Intra}\cap SR_{c,d}^{Inter}$ for both classes of non-commutative supra-Laplacians.
For duplex networks with different intra-layer topologies (the first class), the intra-layer synchronization area is determined by $SR_{c,d}\cap SR_{c,d}^{Intra}$ (as shown by the yellow regions of Figs. 9 and 10).
For duplex networks with nonidentical weighted one-to-one coupling (the second class), the inter-layer synchronization area is determined by $SR_{c,d}\cap SR_{c,d}^{Inter}$,
as  can be seen from  the cyan regions of Figs. 11 and 12. 
These findings shed light on the significant facts that the difference of the intra-layer topologies can lead to the change of the actual inter-layer synchronized regions,
and nonidentical inter-layer one-to-one coupling weights can lead to the change of the actual intra-layer synchronized regions.

In other words, even though here the inter- and intra-layer supra-Laplacian matrices do not commute,
the three synchronized regions still predict the actual areas for complete synchronization and intra-layer synchronization,
or for complete synchronization and inter-layer synchronization.
Therefore, the commutation condition is not necessary for our findings, 
it is only sufficient for our theoretical analysis.
Particularly for the case of different intra-layer topologies,
one can apply these three synchronized regions to predict the actual areas for complete synchronization and intra-layer synchronization.
How generally the observation applies remains an open question.

\section*{Discussion}\label{con}
In summary, 
we develop a master stability function framework
which captures an essential feature of multiplex networks, that the intra-layer and inter-layer coupling functions can be distinct.
Here we define a distinct supra-Laplacian matrix for intra-layer connections, denoted $\mathcal{L}^L$, and one for inter-layer connections, denoted $\mathcal{L}^I$. If $\mathcal{L}^L$ and $\mathcal{L}^I$ commute,
the multiplex network can be easily decoupled and thus the characteristic modes of the intra-layer Laplacian are separated from those of the inter-layer one. (Note this commutation condition is a sufficient but not a necessary condition for our theoretical analysis.)  We can then develop a multiplex master stability equation, Eq.~(\ref{VarEq3}), to establish the  necessary region for complete synchronization.  In the limit of no inter-layer coupling the multiplex MSF reduces to a master stability equation for each independent layer, Eq. (\ref{VarEq4}),  allowing us to calculate the necessary region for intra-layer synchronization.
In the limit of no intra-layer coupling the multiplex MSF reduces to a master stability equation for each independent inter-layer network, Eq. (\ref{VarEq5}),  allowing us to calculate the necessary region  for inter-layer synchronization.

To explicitly use the multiplex MSF framework requires specifying $f(\cdot)$ (i.e., the internal nodal dynamics), and the inter- and intra-layer coupling functions (i.e., $H$ and $\Gamma$ respectively).  We consider specifically a two-layer network of R\"ossler oscillators and various forms of $H$ and $\Gamma$. We find that the different types of coherent behaviors observed in the network are determined by the intersections of the three necessary regions describing complete synchronization, intra-layer synchronization  and inter-layer synchronization.
Given a specified network topology, these regions can then be parameterized by the  intra- and inter-layer  coupling strengths (i.e., $c$ and $d$ respectively).
Complete synchronization is stable when both $c$ and $d$ fall into the overlap of the three regions.
Intra-layer synchronization is stable when both $c$ and $d$ fall into the overlap of the  joint synchronized region and the intra-layer synchronized region.
Inter-layer synchronization is stable when both  $c$ and $d$ fall into the overlap of the  joint synchronized region and the inter-layer synchronized region.


For a given network nodal dynamics, the joint synchronized region is mainly determined by both inner coupling matrices  $H$ and $\Gamma$.
Similarly, the intra-layer synchronized region is mainly determined by the intra-layer coupling matrix $H$,
and the inter-layer synchronized region by the inter-layer coupling matrix $\Gamma$.
Therefore, in addition to nodal dynamics, the inner coupling function is an essential factor to determine which kind of synchronization
the network will arrive at.
If $H$ is in such a form that the intra-layer synchronized region is empty,
intra-layer synchronization is unstable regardless of however large the intra-layer coupling strength is.
Similarly, if 
$\Gamma$ is  in such a form that the inter-layer synchronized region is empty,
inter-layer synchronization is unstable regardless of however large the inter-layer coupling strength is.
In either case, complete synchronization will not occur regardless of the coupling strength.

Here we have  theoretically and numerically  investigated  specific duplex networks of R{\"o}ssler oscillators where the two layers have the same topological structure.  
Our approach can be applied to multiplex networks with different choices for the internal nodal dynamics, different inter- and intra-layer coupling functions, and more layers.  See, for instance the Supplementary Information, Sec 2, for the results of our technique on a three-layer network of R{\"o}ssler oscillators.

As this work introduces a systematic approach for analyzing synchronization patterns in multiplex networks, the focus here is on the simplest case of multiplex networks where the supra-Laplacian matrix of the intra-layer connections is commutative with that of the inter-layer connections. Our framework further holds provided that the multiplex network has intra-layer topology that is identical on each layer and that both the intra-layer Laplacian matrix $L^L$ and the inter-layer Laplacian matrix $L^I$ can be diagonalizable and have real eigenvalues.
We verify numerically in the previous section that the master stability equations derived herein can apply to a broader class of multiplex networks with non-commutative supra-Laplacians, but we can predict only the region of complete synchronization and intra-layer synchronization, or the region of complete synchronization and inter-layer synchronization, and we cannot simultaneously predict these three synchronization behaviors.
Establishing the exact minimal conditions under which our framework can be applied remains an important open question.

\section*{Methods}\label{method}
\noindent\textbf{Decoupling the   multiplex network system.}
Suppose that  supra-Laplacian matrices $\mathcal{L}^L$ and $\mathcal{L}^I$ are symmetric matrices, and satisfy  $\mathcal{L}^L\mathcal{L}^I=\mathcal{L}^I\mathcal{L}^L$,
then there exists an invertible matrix $P$ such that
$$ P^{-1}\mathcal{L}^LP=\mbox{diag}\{\lambda_1,\cdots,\lambda_M,\lambda_{M+1}, \cdots,\lambda_{M\times N}\}, $$
$$ \, P^{-1}\mathcal{L}^IP=\mbox{diag}\{\mu_1,\cdots,\mu_M, \mu_{M+1}, \cdots, \mu_{M\times N}\},$$
where $0=\lambda_1=\cdots=\lambda_M<\lambda_{M+1}\leq\cdots\leq\lambda_{M\times N}$,
$\mu_k\geq 0 \ ~(k=1,2,\cdots,M\times N)$, and $\mbox{diag}\{\upsilon_1,\cdots,\upsilon_M\} $ denotes a diagonal matrix  whose $j$-th diagonal element is $\upsilon_j~( j=1,2, \cdots, M).$

By denoting a new vector $\bm{\eta}=[\bm{\eta}_1^\top, \bm{\eta}_2^\top, \cdots, \bm{\eta}_{{M\times N}}^\top]^\top=(P\otimes I_m)^{-1} \bm{\xi}$, we can turn the variational equation (\ref{VarEq0})  into
\begin{equation}\label{VarEq1}
  \dot{\bm{\eta}}=[I_{M\times N}\otimes Df(\bm{s})-c(\mbox{diag}\{\lambda_1,\cdots,\lambda_{M\times N}\}\otimes H)
    -d(\mbox{diag}\{\mu_1,\cdots,\mu_{M\times N}\}\otimes\Gamma)]\bm{\eta}.
\end{equation}
It further yields
\begin{equation}\label{VarEq2}
  \dot{\bm{\eta}_k}=[Df(\bm{s})-c\lambda_k H
    -d\mu_k\Gamma]\bm{\eta}_k,\quad k=1,2,\cdots,M\times N.
\end{equation}
Here, $\bm{\eta}_k$  represents the mode of perturbation in the generalized eigenspace associated with $\lambda_k$ and $\mu_k$.
A criterion for the synchronization manifold to be (asymptotically) stable is
that all the transversal Lyapunov exponents of the variational equation (\ref{VarEq2})   are strictly
negative.  Clearly, these Lyapunov exponents depend on the node dynamics  $f(\cdot)$,
the network intra- and inter-layer coupling strengths $c$ and $d$,  and the coupling matrices $H$ and $\Gamma$.
Consequently, we can get the three master stability equations: Eqs. (\ref{VarEq3}),   (\ref{VarEq4}) and  (\ref{VarEq5}).

\vspace{1cm}
\noindent\textbf{Calculating synchronized regions $\mathbf{SR_{\,c,d}}$.}
We can calculate three synchronized regions with regard to parameters $\alpha$ and $\beta$:
$SR$, $SR ^{Intra}$ and $SR^{Inter}$ from Eqs. (\ref{VarEq3}),  (\ref{VarEq4}) and   (\ref{VarEq5}), respectively.
Furthermore, when the network topologies are given, we can directly calculate the characteristic values of supra-Laplacian matrices  and parameterize those regions in terms of $c$ and $d$, since  $\alpha=c\lambda $ and $\beta=d\mu$.

{ For example, when $H=I_{11}$ and $\Gamma=I_{11}$, the nonzero characteristic modes
$\alpha=c\lambda$ and $\beta=c\mu $ should lie in $SR=\{(\alpha,\beta)|\,0.2<\alpha+\beta<4.6\}$,
and consequently the region with respect to parameters  $c$ and $d$ is
$$SR_{\,c,d}=\{(c, d)|\,0.2<c+2d, c+0.4d<0.92\}.$$ }
For other combinations of $H$ and $\Gamma$, the synchronized regions with respect to parameters $c$ and $d$ can be similarly obtained.

\vspace{1cm}
\noindent\textbf{Synchronization errors \& Indicator function.}
To measure  the extent of intra-layer,  inter-layer and complete synchronization,
we introduce the following indices:
\begin{equation}
  E_{Intra}^{(k)}(t)=\frac{1}{N}\sum_{i=1}^N\|x_i^{(k)}(t)-\overline{x}^{(k)}(t)\|,\; k=1,2,\cdots,M
\end{equation}
where $\|\cdot\|$ is a norm operator, and $\overline{x}^{(k)}(t)$ is the average state of all the nodes in the $k$th layer  at time $t$.
 Thus $E_{Intra}^{(k)}(t)$ is the synchronization error of nodes in the $k$th layer at time $t$, namely,  the intra-layer synchronization error.

Similarly, the inter-layer synchronization error is defined as
\begin{equation}
  E_{Inter}(t)=\frac{1}{MN}\sum_{i=1}^N\sum_{k=1}^M\|x_i^{(k)}(t)-\overline{x}_i(t)\|,
\end{equation}
and  the complete synchronization error is defined as
\begin{equation}
  E(t)=\frac{1}{NM}\sum_{k=1}^M\sum_{i=1}^N\|x_i^{(k)}(t)-\overline{x}(t)\|,
\end{equation}
where $\overline{x}_i(t)$ is the average state of the node $i$ in each layer and its counterparts in other layers,
and $\overline{x}(t)$ is that of all the nodes in the multiplex network.

With these definitions, we use the following indicator function to represent complete synchronization, intra-layer synchronization and inter-layer synchronization: 
\begin{equation}\label{indicator}
       Id=\left\{\begin{array}{l}
                     3, \;E_{Inter}(t)<\epsilon\; \text{and}\; E_{Intra}^{(k)}(t)<\epsilon \; \text{for all}\; t>T_0,\\
                     2, \;E_{Inter}(t)\geq\epsilon\; \text{and}\; E_{Intra}^{(k)}(t)<\epsilon\; \text{for all}\; t>T_0,\\
                     1, \;E_{Inter}(t)<\epsilon\; \text{and}\; E_{Intra}^{(k)}(t)\geq\epsilon\; \text{for all}\; t>T_0,\\
                     0, \; \text{other}.
                   \end{array}\right.
\end{equation}
Here, $T_0$ is a time threshold value and $\epsilon$ is a given threshold for synchronization errors.
In the simulations, $\epsilon=1.0\times10^{-2}$, and $T_0=0.8T_{total}$ ($T_{total}$ is the total evolution time).
It is obvious  that the network reaches complete synchronization when $Id=3$,
intra-layer synchronization when $Id=2$, inter-layer synchronization when $Id=1$, and none of the above when $Id=0$.


\section*{Acknowledgments}
This work is supported in part by the National Key Research and Development Program of China under Grant 2016YFB0800401,
in part by the National Natural Science Foundation of China under Grants 61573004, 11501221, 61573262, 61621003 and 61532020,
in part by the U.S. Army Research Office under Multidisciplinary University Research Initiative Award No. W911NF-13-1-0340 and Cooperative Agreement W911NF-09-2-0053, in part by DARPA grant W911NF-17-1-0077,
in part  by the Promotion Program for Young and Middle-aged Teacher in Science and Technology Research of Huaqiao University (ZQN-YX301),
and in part by the Natural Science Foundation of Fujian Province (2015J01260).

\end{document}